\documentclass[11pt]{article}  
\usepackage{graphicx,floatflt,amssymb,epsf}   
\begin{document}  
\begin{flushright}  
SINP/TNP/07-28\\  
\end{flushright}  
  
\vskip 5pt  
  
\begin{center}  
{\Large \bf CP Violation and Flavour Mixings in Orbifold GUTs} \\  
\vspace*{1cm}  
\renewcommand{\thefootnote}{\fnsymbol{footnote}}  
{\large {\sf Gautam Bhattacharyya,${}^{1,}$\footnote{E-mail 
 address: {\tt gautam.bhattacharyya@saha.ac.in}}}
{\sf Gustavo C.~Branco,${}^{2,3,}$\footnote{E-mail 
 address: {\tt gbranco@mail.ist.utl.pt}}}  and        
{\sf Joaquim I.~Silva-Marcos${}^{2,}$\footnote{E-mail 
 address: {\tt joaquim.silva-marcos@cern.ch}}}  
} \\  
\vspace{10pt}  
{\small ${}^{1)}${\em Saha Institute of Nuclear Physics,  
1/AF Bidhan Nagar, Kolkata 700064, India} \\ 
${}^{2)}${\em 
Departamento de F\'{\i}sica and Centro de 
F\'{\i}sica Te\'orica de Part\'{\i}culas (CFTP), \\Instituto Superior 
T\'{e}cnico, Av. Rovisco Pais, P-1049-001 Lisboa, Portugal } \\
${}^{3)}${\em 
Physik-Department, Technische Universit\"at M\"unchen,
James-Franck-Strasse,\\ D-85748 Garching, Germany}
}   
  
\normalsize  
\end{center}  
  
\begin{abstract}  

We address the flavour problem by incorporating the hypothesis of
universal strength of Yukawa couplings in the framework of a 5D GUT
model compactified on an $S^1/(Z_2 \times Z_2^{\prime})$ orbifold. We
show that a quantitatively successful picture of fermion masses and
mixings emerges from the interplay between the bulk suppression factors
of geometric origin and the phases of the Yukawa matrices. We give an
explicit example, where we obtain a good fit for both the CKM and PMNS
matrices.

%\vskip 5pt \noindent  
%\texttt{PACS Nos:~11.10.Kk, 11.30.Hv} \\ 
%\texttt{Key Words:~CP violation, Orbifold compactification}  
\end{abstract}

\renewcommand{\thesection}{\Roman{section}}  
\setcounter{footnote}{0}  
\renewcommand{\thefootnote}{\arabic{footnote}}

\section{Introduction and motivation}

The origin of the observed pattern of fermion masses and flavour mixings is
a central and open question in particle physics. In an attempt to seek
further insight into the flavour problem, we propose a framework which links
two main hypotheses:

(i) The SU(5) gauge symmetry of a 5-dimensional (5D) Grand Unified
Theory (GUT) is broken by orbifold boundary conditions. Some matter
multiplets are confined at the two orbifold fixed points, while others
access the bulk \cite{prev-1,prev-2,br}.  The scalar multiplets which
acquire vacuum expectation values (VEVs) are in the bulk.

(ii) Apart from the suppression factors of Yukawa couplings determined
solely by the relative locations of the fermions, there is a Universal
Strength of Yukawa couplings (USY) \cite{usy}. However, in order to
incorporate CP violation, we introduce complex Yukawa couplings with
flavour dependent phases.

In this paper, we elaborate on these two ideas in some detail, and
illustrate how they lead to a qualitative and quantitative understanding of
fermion masses and mixings.

Higher dimensional orbifold GUTs possess some advantages over the
conventional ones, such as providing a simpler realization of
doublet-triplet splitting which is possible as a consequence of GUT gauge
symmetry breaking by boundary conditions. Embedding supersymmetry in such a
higher dimensional GUT further helps in achieving gauge coupling
unification. Following the initial attempts \cite{prev-1}, a series of
investigations \cite{prev-2} has been launched aiming at constructing a
supersymmetric SU(5) gauge theory using an $S^{1}/(Z_{2}\times Z_{2}^{\prime
})$ orbifold. Such scenarios allow for a novel handling of the flavour
problem by assuming that flavour hierarchies may arise from a suppression
mechanism of geometrical origin, where the suppression depends on the
relative location of the fields in the extra dimension ($y$). The physical
region stretches on a quarter-circle $y=[0,\pi R/2]$, where $R$ is the
radius of compactification. Our matter fields consist of SU(5) multiplets of
different generations ($\mathbf{\bar{5}_i}$, $\mathbf{{10}_i}$), and the
right-handed singlet neutrinos ($\mathbf{{N}_i}$). The orbifold allows three
locations where these matter fields can be placed: the SU(5) preserving 5D
bulk, and the two fixed points at $y=0$ (the $O$ brane) and at $y=\pi R/2$
(the $O^{\prime}$ brane). The $O$ brane has the full SU(5) symmetry, while
at the $O^{\prime}$ brane the unbroken gauge symmetry is that of the
Standard Model (SM). If a given matter field spreads into the bulk, its 4D
zero mode has a wavefunction normalization factor, given by $\epsilon$ (see
the next section for details), attached to its Yukawa coupling. This is in
fact a suppression factor (typically, $\epsilon \sim 0.1$) of pure geometric
origin. On the other hand, there is no such suppression when the concerned
field is localized at one of the two branes. Thus, besides an overall Yukawa
strength, the entries of the mass (or, Yukawa) matrices are either 1 or
powers of $\epsilon$ depending on the relative localization of the matter
fields.

An important feature of the previous analyses \cite{prev-2} is that
all Yukawa couplings have been assumed to be real, thus neglecting CP
violation.  Moreover, for generating quantitatively successful
Cabibbo-Kobayashi-Maskawa (CKM) and Pontecorvo-Maki-Nakagawa-Sakata
(PMNS) matrices, one further needed to exploit the ${\mathcal{O}}(1)$
uncertainties of the coefficients of $\epsilon$-dependent terms. In
fact, it has been shown that with real Yukawa couplings, or even with
complex Yukawa couplings where the phases are also decided by the
location of the fields, unless one exploits these ${\mathcal{O}}(1)$
uncertainties, it is not possible to simultaneously satisfy even the
gross qualitative features of the CKM and PMNS matrices, no matter
what locations are chosen for the matter multiplets \cite{br}. The
introduction of these uncertainties, which cannot be calculated, adds
a gross arbitrariness to the framework. It is not in fact surprising
that their exact values have not been spelt out in the existing
literature \cite{prev-2}.

Clearly, the whole framework would be more appealing if the
qualitative and quantitative features of both quark and lepton masses
and mixings could be reproduced without having to make use of those
uncertainties. This constitutes the main motivation of the present
paper, where we attempt to generate a quantitatively successful
pattern of fermion masses and mixings, implementing the USY hypothesis
in the orbifold GUT scenarios. We take into account the phases
associated with USY and the volume suppression factors related
to orbifolding, without using the above mentioned ${\mathcal{O}}(1)$
uncertainties.

Another motivation for the present work stems from recent results from
experiment. \emph{First}, after the recent measurement of the angle
$\gamma $ of the unitarity triangle, it is now
mandatory to consider complex quark mass matrices leading to a
\emph{complex} CKM matrix. Prior to this measurement, one could have
had a real CKM matrix and still explain the observed CP violation in
the $K$ and $B$ systems by invoking new physics with non-standard
phases contributing via loops to the $K^{0}-\overline{K^{0}}$ and
$B^{0}-\overline{B^{0}}$ mixings. However, the extraction of $\gamma$
from experimental data involves tree level process where new physics
is unlikely to compete with the SM contributions. Thus the measurement
of $\gamma $ provides an irrefutable evidence for a complex CKM matrix
\cite{gamma}, which in the present framework implies complex Yukawa
couplings. \emph{Second}, although the USY hypothesis in the SM has
been shown to reproduce the correct pattern of quark masses and
mixings, it leads in general to a rather small value for the strength
of CP violation, measured by the imaginary part of any
rephasing-invariant CKM matrix quartet. We will show that when
incorporated in a higher dimensional orbifold framework, the USY
hypothesis can lead to a sufficiently large strength of CP violation.

\section{Formalism of GUT orbifolding}

Our formalism is based on a 5D GUT scenario with $N=1$ supersymmetry
where the gauge symmetry is SU(5). The extra dimension ($y$) is
compactified on an orbifold $S^{1}/(Z_{2}\times Z_{2}^{\prime })$, and
the inverse radius $ R^{-1}$ is chosen to be of the order of $10^{16}$
GeV. $S^{1}$ is first divided by $Z_{2}$ ($y\leftrightarrow -y$) and
then further by $ Z_{2}^{\prime }$ ($y^{\prime }\leftrightarrow
-y^{\prime }$ with $y^{\prime }=y+\pi R/2$), restricting the physical
spacetime within the interval $ [0,\pi R/2]$. The $O$ and $O^{\prime
}$ branes are located at the fixed points $y=0$ and $y=\pi R/2$
respectively. Now, let us consider a generic 5D bulk field $
\phi (x^{\mu },y)$. The $Z_{2}$ and $Z_{2}^{\prime }$ parities (called $P$
and $P^{\prime }$, respectively) act on this field as 
\begin{eqnarray}
\phi (x^{\mu },y) &\rightarrow &\phi (x^{\mu },-y)=P\phi (x^{\mu },y), 
\nonumber \\
\phi (x^{\mu },y^{\prime }) &\rightarrow &\phi (x^{\mu },-y^{\prime
})=P^{\prime }\phi (x^{\mu },y^{\prime }).
\end{eqnarray}
The 4D Kaluza-Klein (KK) towers with $(P,P^{\prime })=(\pm ,\pm )$ are the
following: $\phi _{++}^{(2n)}$ with a mass $2n/R$, $\phi _{+-}^{(2n+1)}$ and 
$\phi _{-+}^{(2n+1)}$ with a mass $(2n+1)/R$ and $\phi _{--}^{(2n+2)}$ with
a mass $(2n+2)/R$. Thus only $\phi _{++}^{(2n)}$ have massless modes. Also,
only $\phi _{++}$ and $\phi _{+-}$ have non-vanishing components at the $O$
brane. From a 4D perspective we have two different $N=1$ supersymmetries.
Now assigning suitable $(P,P^{\prime })$ quantum numbers to the fields, it
is possible to project out one $N=1$ supersymmetry leaving the other
unbroken.

The choice of $P=(+++++)$ and $P^{\prime }=(---++)$ or $(+++--)$ acting on a 
\textbf{5} keeps SU(5) intact at $O$ but breaks SU(5) to $\mathrm{SU(3)}_{
\mathrm{C}}\otimes \mathrm{SU(2)}_{\mathrm{L}}\otimes \mathrm{U(1)}_{\mathrm{
Y}}$~ at $O^{\prime }$. With the above $P^{\prime }$ assignments it is
not possible to fill up a complete SU(5) multiplet with zero modes. We
need to introduce $\mathbf{\bar{5}^{\prime }}$ and $\mathbf{10^{\prime
}}$ with $ P^{\prime }$ assignments opposite to those in
$\mathbf{\bar{5}}$ and $\mathbf{10}$ in order to obtain the correct
low energy matter content.

We include all Yukawa couplings that are consistent with gauge
symmetry and R-parity. We assume that the Higgs multiplets always
reside in the bulk. If we denote a Yukawa coupling involving three
brane superfields by $\lambda $, then the one where one of the three
fields is a bulk field is $\lambda /\sqrt{M_{\ast }R}$, where $M_{\ast
}$ is the ultraviolet cutoff of the 5D theory. The appearance of
$M_{\ast }$ is related to the canonical normalization of the zero mode
kinetic terms. It has been shown in Ref.~\cite{prev-2} that
$(M_{\ast}R)\sim 10^{2-3}$ is a good choice for gauge coupling
unification. So $\epsilon =1/\sqrt{M_{\ast }R}\sim 0.1$ acts as a
suppression factor. The number of such factors multiplying $\lambda $
is given by the number of bulk zero modes in a given interaction, each
bulk field contributing with one $\epsilon $.

\section{Quark and lepton mass matrices}

We write the fermion mass matrices in the convention that the fields
on the left are left-handed and those on the right are
right-handed. The up quark mass terms are given by
$\overline{\mathbf{10}}_{i}(M_{u})_{ij}\mathbf{10^{c}}_{j}$, with
$M_{u}$ symmetric. The down quark mass terms are written as
$\overline{\mathbf{10}}_{i}(M_{d})_{ij}\mathbf{5^{c}}_{j}$, while the
charged lepton mass matrix $M_{l}$ is simply $(M_{d})^{\dagger }$. The
effective light neutrino Majorana mass matrix terms are given by
$\bar{\mathbf{5}}_{i}(M_{\nu }^{\mathrm{Maj}})_{ij}
\bar{\mathbf{5}}_{j}$,
after integrating out the $\mathbf{N_{i}}$ states.

In order to obtain the flavour structure of the fermion mass matrices,
we have to specify the matter multiplets of the different
generations. Although there is no strict criteria in this regard, let
us be guided by some phenomenological considerations: (i) we place
$\mathbf{\bar{5}_{1}}$ and $\mathbf{{10}_{1}}$ in SU(5) bulk, then the
first generation zero mode quarks and leptons come from different
SU(5) multiplets, which prevents proton decay to occur at leading
order; (ii) we place $\mathbf{\bar{5}_{3}}$ and $\mathbf{{10}_{3}}$
at the SU(5) preserving $O$ brane, which helps to ensure
$m_{b}=m_{\tau }$ at the GUT scale; (iii) we choose to put
$\mathbf{\bar{5}_{2}}$ in the bulk and $\mathbf{{10}_{2}}$ at the $O$
brane; finally (iv) all $\mathbf{N}_{i}$ are placed in the bulk, as it
is difficult to confine on a brane a particle which does not have any
gauge charge.

With the above assignments, we can write the different mass matrices as: 
\begin{equation}  \label{mud}
M_{u}=\lambda _{u}\ \left[ 
\begin{array}{ccc}
\epsilon _{u}^{2} & \epsilon _{u} & \epsilon _{u} \\ 
\epsilon _{u} & 1 & e^{ia_{u}} \\ 
\epsilon _{u} & e^{ia_{u}} & e^{ib_{u}}
\end{array}
\right] \quad ;\quad M_{d}=\lambda _{d}\ \left[ 
\begin{array}{ccc}
\epsilon _{d}^{2} & \epsilon _{d}^{2} & \epsilon _{d} \\ 
\epsilon _{d} & \epsilon _{d}\ e^{ia_{d}} & 1 \\ 
\epsilon _{d} & \epsilon _{d}\ e^{ib_{d}} & e^{ic_{d}}
\end{array}
\right] \quad ;\quad  \label{quarkmass}
\end{equation}
\begin{equation}  \label{nul}
M_{\nu }=\lambda _{\nu }\ \left[ 
\begin{array}{ccc}
\epsilon _{\nu }^{2} & \epsilon _{\nu }^{2}\ e^{ia_{\nu }} & -\epsilon _{\nu
} \\ 
\epsilon _{\nu }^{2}\ e^{ia_{\nu }} & \epsilon _{\nu }^{2}\ e^{ib_{\nu }} & 
-\epsilon _{\nu } \\ 
-\epsilon _{\nu } & -\epsilon _{\nu } & 1
\end{array}
\right] \quad ;\quad M_{l}=\rho \ M_{d}^{\dagger }\quad .  
\label{leptonmass}
\end{equation}

A few comments are in order. As shown in \cite{usy}, within USY, in
general in each charged sector, only four phases contribute to the
spectrum of fermion masses. By making appropriate weak-basis
transformations, one can make some changes in the location of the
phases, without altering the physical consequences of the model. In
Eq.~(\ref{quarkmass}) we have just made one such convenient choice.
Since the $\epsilon_f$'s are by definition positive, the negative
signs in some entries in the neutrino mass matrix correspond to a
phase choice of $\pi$. In this paper we do not scan the entire set of
phase choices, but just provide an example which leads to the right
mixing and mass spectrum, both in the quark and lepton sectors. The
geometric suppression factors and the USY phases are in general
flavour dependent. The factor $\rho $ appearing before the charged
lepton mass matrix deserves some mention. We recall that zero mode
bulk matter fields of a given generation arise from different SU(5)
multiplets. Yet, for simplicity and predictivity, we maintain the
ususal relationship between $M_l$ and $M_d^\dagger$ modulo an overall
$\rho$ factor.  The deviation of $\rho$ from unity may accrue due to
induced effects from the $O^{\prime}$ brane. Renormalisation group
running effects too may partly account for such deviation, although a
detailed study is beyond the scope of the present paper.  From a more
practical point of view, for numerical fitting of the charged lepton
mass matrix, we require $\rho \sim 0.6$.

Notice that each of these matrices is of type $M=\lambda \ D_{1}\
M^{\mathrm{ USY}}\ D_{2}$, where $D_{1,2}=$ diag $(\epsilon
^{n},\epsilon ^{m},1)$ for $ n,m=0,1$. The matrix $M^{\mathrm{USY}}
=(e^{ia_{ij}})$ has a USY-type texture, i.e. all Yukawa couplings in
a given matrix have identical moduli but differ in some complex
phases. As an example, one can write the down quark mass matrix 
of Eq.~(\ref{quarkmass}) as
\begin{equation}
M_{d}=\lambda _{d}\ \mathrm{diag}(\epsilon _{d},1,1)\ \left[ 
\begin{array}{ccc}
1 & 1 & 1 \\ 
1 & e^{ia_{d}} & 1 \\ 
1 & e^{ib_{d}} & e^{ic_{d}}
\end{array}
\right] \ \mathrm{diag}(\epsilon _{d},\epsilon _{d},1)\quad .
\label{massa-u}
\end{equation}

To study the parameter space, it is instructive to calculate some invariants
of $H_{f}=M_{f}M_{f}^{\dagger }$. We first define the dimensionless
matrix 
$H^{\prime}_f\equiv H_f/t$, where $t=\mathrm{Tr}(H_f)$. Then noting that 
\begin{equation}
\det \left( H^{\prime}_f\right) =\frac{\left( \frac{m_{1}}{m_{3}}\right)
^{2}\left( \frac{m_{2}}{m_{3}}\right) ^{2}}{\left( 1+\left( \frac{m_{1}}{
m_{3}}\right) ^{2}+\left( \frac{m_{2}}{m_{3}}\right) ^{2}\right) ^{3}}\quad ,
\label{relation}
\end{equation}
and using the known fermion mass hierarchies, one obtains in leading order
of $\epsilon _{u}$, $\epsilon _{d}$ and $\epsilon _{\nu }$ ~: 
\begin{equation}
\begin{array}{lll}
\frac{1}{2}\epsilon _{u}^{2}\sin ^{2}(\frac{a_{u}}{2})=\left( \frac{m_{u}}{
m_{t}}\right) \left( \frac{m_{c}}{m_{t}}\right) & \quad ;\quad & \sqrt{2}
\epsilon _{d}^{3}|\sin (\frac{a_{d}}{2})\sin (\frac{c_{d}}{2})|\simeq \left( 
\frac{m_{d}}{m_{b}}\right) \left( \frac{m_{s}}{m_{b}}\right) \quad ; \\ 
&  &  \\ 
4\epsilon _{\nu }^{4}\sin ^{2}(\frac{a_{\nu }}{2})=\frac{\left( \frac{m_{\nu
1}}{m\nu _{3}}\right) \left( \frac{m_{\nu 2}}{m_{\nu 3}}\right) }{\left(
1+\left( \frac{m_{\nu 1}}{m_{\nu 3}}\right) ^{2}+\left( \frac{m_{\nu 2}}{
m_{\nu 3}}\right) ^{2}\right) ^{\frac{3}{2}}} & \quad ;\quad & 
\end{array}
\label{leading}
\end{equation}
where (except for the neutrinos) we kept only the leading order terms in the
masses. Just to have a numerical feel, taking e.g. $\epsilon _{u}=0.1$ leads
to $|a_{u}|=(4-9)\times 10^{-3}$. Now we construct the second invariant of $
H^{\prime}_f$, given by $\chi (H^{\prime}_f)=\lambda _{1}\lambda _{2}+\lambda
_{2}\lambda _{3}+\lambda _{3}\lambda _{1}$, where $\lambda _{i}$'s denote
the $H^{\prime}_f$ eigenvalues. In terms of fermion mass ratios,  
\begin{equation}  \label{chidef}
\chi \left( H^{\prime }\right) =\left( \frac{m_{2}}{m_{3}}\right) ^{2}\frac{
1+\left( \frac{m_{1}}{m_{2}}\right) ^{2}+\left( \frac{m_{1}}{m_{3}}\right)
^{2}}{\left( 1+\left( \frac{m_{1}}{m_{3}}\right) ^{2}+\left( \frac{m_{2}}{
m_{3}}\right) ^{2}\right) ^{2}}\quad .  \label{relation1}
\end{equation}
Using Eqs.~(\ref{mud}), (\ref{nul}) and (\ref{chidef}), one obtains
the following relations in leading order of $\epsilon _{u}$, $\epsilon _{d}$
and $\epsilon _{\nu }$: 
\begin{equation}
\begin{array}{l}
\chi \left( H_{u}^{\prime }\right) =\frac{1}{4}\sin ^{2}(a_{u}-\frac{b_{u}}{2
})=\left( \frac{m_{c}}{m_{t}}\right) ^{2}\quad ; \\ 
\chi \left( H_{d}^{\prime }\right) =\epsilon _{d}^{2}\left[\sin^{2}(\frac{
c_{d}}{2})+\sin ^{2}(\frac{a_{d}+c_{d}-b_{d}}{2})\right] =\left( \frac{m_{s}
}{m_{b}}\right) ^{2} \quad ; \\ 
\chi \left( H_{\nu }^{\prime }\right) =4\epsilon _{\nu }^{2}
\left[2\sin ^{2}(\frac{a_{\nu }}{2})+\sin ^{2}(\frac{b_{\nu }}{2})\right]=
\left( \frac{m_{\nu 2}}{m_{\nu 3}}\right) ^{2}
\frac{1+\left( \frac{m_{\nu 1}}{m_{\nu 2}}\right)^{2} +
\left( \frac{m_{\nu 1}}{m_{\nu 3}}\right) ^{2}}{\left( 1+\left( \frac{
m_{\nu 1}}{m_{\nu 3}}\right) ^{2}+\left( \frac{m_{\nu 2}}{m_{\nu 3}}\right)
^{2}\right) ^{2}}\quad ;
\end{array}
\label{chi}
\end{equation}
where we have kept only the leading order of the mass ratios of the quarks
and charged leptons, while for the neutrinos the right-hand side is exact.

The above matrices can be diagonalized through: 
\begin{equation}
\begin{array}{lll}
V_{u}^{\dagger }\ M_{u}M_{u}^{\dagger }\ V_{u}=D_{u}^{2}\quad , &  & 
V_{d}^{\dagger }\ M_{d}M_{d}^{\dagger }\ V_{d}=D_{d}^{2}\quad , \\ 
&  &  \\ 
V_{l}^{\dagger }\ M_{l}M_{l}^{\dagger }\ V_{l}=D_{l}^{2}\quad , &  & V_{\nu
}^{\mathrm{T}}\ M_{v}^{\mathrm{Maj}}\ V_{v}=D_{\nu }\quad ,
\end{array}
\label{diag}
\end{equation}
where the $D_{f}$ are diagonal matrices containing the masses of the
fermions for each $f=u,d,l,\nu $. Notice that the diagonalization of the
neutrino mass matrix is somewhat different, because it is a symmetric matrix
obtained after the integration of the bulk $\mathbf{N_{i}}$ states. The
physical mixing matrices in the quark and lepton sectors are given by:
\begin{equation}
V_{\mathrm{CKM}}=V_{u}^{\dagger }V_{d}\quad ;\quad V_{\mathrm{PMNS}
}=V_{l}^{\dagger }V_{\nu }\quad .  \label{ckmns}
\end{equation}

In the next section, we provide an explicit example where the correct
spectrum of fermion masses and mixing is obtained, including $J=\mathrm{Im}
(V_{us}V_{cb}V_{ub}^{\ast}V_{cs}^{\ast})$, which controls the strength of
CP violation in the quark sector. In order to obtain a sufficiently large
value of $J$, it is crucial to have $\epsilon _{d}$ not exactly equal to $
\epsilon _{u}$. To understand the requirement $\epsilon _{d}\neq \epsilon
_{u}$, let us first recall the identity Tr $([H_{u},H_{d}]^{3})= 6\
\Delta \ J$, where $\Delta =
(m_{t}^{2}-m_{u}^{2})(m_{t}^{2}-m_{c}^{2})
(m_{c}^{2}-m_{u}^{2})(m_{b}^{2}-m_{d}^{2})(m_{b}^{2}-m_{s}^{2})
(m_{s}^{2}-m_{d}^{2})$. For simplicity of presentation, we take the
limit $m_{u}=0$ setting $a_{u}=0$. If we now take $\epsilon
_{d}=\epsilon _{u}=\epsilon $, and employ Eqs.~(\ref{leading}) and
(\ref{chi}) we obtain
\begin{equation}
|J|\ \approx \frac{1}{\sqrt{2}}\ \epsilon ^{5}\ \left( \frac{m_{d}}{m_{s}}
\right) \left\vert \sin (\frac{a_{d}}{2})\right\vert \left\vert 2\sin ^{2}(
\frac{a_{d}}{2})-\sin ^{2}(\frac{a_{d}-b_{d}}{2})\right\vert \ . 
\label{upperbound}
\end{equation}
Putting $\epsilon =0.1$ leads to $|J|\ \approx 10^{-7}$, which is too small.
On the other hand, if $\epsilon _{d}\neq $ $\epsilon _{u}$, one finds in
leading order, 
\begin{equation}
\left\vert J\right\vert \ \approx \frac{1}{\sqrt{2}}\ |\epsilon
_{d}-\epsilon _{u}|\ \left( \frac{m_{d}}{m_{s}}\right) \left\vert \sin (
\frac{a_{d}-b_{d}}{2})\cos (\frac{b_{d}}{2})\ \right\vert   \label{Jagain}
\end{equation}
which for $|\epsilon _{d}-\epsilon _{u}|=4\times 10^{-3}$ can be as large as 
$10^{-5}$. We checked that these estimates are very close to the exact
numbers obtained without any approximation. 

We recall that, in the conventional 4D USY case, $|J|$ is quite small
because all phases have to be small due to the strong quark mass
hierarchy and due to the smallness of $|V_{ub}|$ and $|V_{cb}|$. Here,
with the embedding in a 5D orbifold GUT, some of the phases, e.g. in
the down quark sector, may be\ quite large because the mass hierarchy
is now ensured through powers of $\epsilon $ entering the mass ratio
relations, as can be seen from Eqs.~(\ref {leading}) and
(\ref{chi}). This is essentially the main reason why in the present model a
sufficiently large value of $|J|$ can be obtained. As we have just
seen, the effect is magnified even by a slight variation
from the equality between $\epsilon_d$ and $\epsilon_u$, namely
$|\epsilon_d - \epsilon_u|/(\epsilon_d + \epsilon_u) = {\cal{O}}
(1\%)$.

\section{An explicit numerical example}

In this example, the charged lepton quark matrix $M_{l}$ has the same
structure as the Hermitian conjugate of the down quark mass matrix $M_{d}$.
However, for a good numerical fit, we have to choose different values
for the phases. This is to avoid exact mass ratio relations between the
charged leptons and down quarks. Thus we choose,
\begin{equation}
M_{l}=\rho \left( M_{d}^{\prime }\right) ^{\dagger }  \label{diflepton}
\end{equation}
where
\[
M_{d}^{\prime }=\left[ 
\begin{array}{ccc}
\epsilon _{d}^{2} & \epsilon _{d}^{2} & \epsilon _{d} \\ 
\epsilon _{d} & \epsilon _{d}\ e^{ia_{d}^{\prime }} & 1 \\ 
\epsilon _{d} & \epsilon _{d}\ e^{ib_{d}^{\prime }} & e^{ic_{d}^{\prime }}
\end{array}
\right] .
\]

Using Eqs.~(\ref{quarkmass}), (\ref{leptonmass}) and (\ref{diflepton})
and with the input
\begin{equation}
\begin{array}{lllll}
\lambda _{u}=89.55\ {\rm GeV} \quad & \epsilon _{u}=0.1 \quad & 
a_{u}=-0.00441 \quad & b_{u}=0.007 &  \\ 
\lambda _{d}=2.09\ {\rm GeV} \quad & \epsilon _{d}=0.1046 \quad 
& a_{d}=0.87
\quad & b_{d}=0.495 \quad & c_{d}=0.084 \ ; \\ 
\rho =0.59  & \quad & a_{d}^{\prime }=0.06 \quad & b_{d}^{\prime }=-0.1
\quad & c_{d}^{\prime }=0.8  \\ 
\lambda _{v}=0.0387\ {\rm eV} \quad & \epsilon _{_{\nu }}=0.45
 \quad  & a_{\nu
}=0.2 \quad & b_{\nu }=0.43   & 
\end{array}
\label{values}
\end{equation}
one obtains the following output for the quark masses (at the weak scale): 
\begin{equation}
\begin{array}{l}
m_{u}=1.1\ {\rm MeV} \\ 
m_{c}=709\ {\rm MeV} \\ 
m_{t}=180\ {\rm GeV}
\end{array}
\quad ;\quad 
\begin{array}{l}
m_{d}=3.42\ {\rm MeV} \\ 
m_{s}=72.4\ {\rm MeV} \\ 
m_{b}=3.0\ {\rm GeV}
\end{array} ;
\label{quarks}
\end{equation}
CKM mixing :
\begin{equation}
\left\vert V_{\rm CKM}\right\vert =\left[ 
\begin{array}{ccc}
0.9755 & 0.2201 & 0.0039 \\ 
0.2199 & 0.9748 & 0.0374 \\ 
0.0102 & 0.0362 & 0.9993
\end{array} 
\right] \quad ;\quad 
\begin{array}{l}
\left\vert \frac{V_{ub}}{V_{cb}}\right\vert =0.104 \\ 
|J|=2.91\times 10^{-5} \\ 
\sin (2\beta )=0.65
\end{array} ; 
\label{ckm}
\end{equation}
charged lepton and neutrino masses :
\begin{equation}
\begin{array}{l}
m_{e}=0.50\ {\rm MeV} \\ 
m_{\mu }=105.2\ {\rm MeV} \\ 
m_{\tau }=1.77\ {\rm GeV}
\end{array}
\quad ;\quad 
\begin{array}{l}
\Delta m_{21}^{2}=7.09\times 10^{-5}\ {\rm eV}^{2} \\ 
\Delta m_{31}^{2}=2.89\times 10^{-3}\ {\rm eV}^{2} \\ 
m_{\nu 3}=0.0504\ {\rm eV}
\end{array} ;
\quad  \label{leptons}
\end{equation}
and PMNS mixing : 
\begin{equation}
\begin{array}{l}
\left\vert V_{\rm PMNS}\right\vert =\left[ 
\begin{array}{ccc}
0.8379 & 0.5396 & 0.0821 \\ 
0.3925 & 0.6759 & 0.6337 \\ 
0.3792 & 0.5020 & 0.7773
\end{array}
\right] \quad ;\quad 
\begin{array}{l}
\sin ^{2}(\theta _{12})=0.293 \\ 
\sin ^{2}(\theta _{23})=0.392 \quad ;\quad |J'|=0.0126\\ 
|U_{13}|^{2}=0.0067
\end{array} ,
\end{array}
\label{mns}
\end{equation}
where $J'$ measures the strength of CP violation in the lepton sector,
which is expected to be large because this sector involves two large
mixing angles.

\section{Conclusions}

We have shown that the ansatz of universality of strength of Yukawa
couplings, when implemented on a 5D orbifold GUT model, can
accommodate the observed pattern of quark and lepton masses and
mixings. This is achieved by taking into account the geometric
suppression factors, arising from the relative locations of the
fermion fields, which appear as powers of $\epsilon$, and allowing for
complex phases in the Yukawa couplings. We reiterate that in view of
recent experimental data, the CKM matrix is necessarily complex, even
if one allows for the presence of physics beyond the SM. This renders
mandatory the introduction of complex Yukawa couplings which, apart
from the geometric suppression factors, have all the same modulus
within the framework considered. It is remarkable that a good fit of
the fermion masses and mixings is obtained, without having to invoke
the order one uncertainties in the moduli of the Yukawa matrix elements.

Another important point is the fact that embedding USY in a 5D
orbifold GUT enables one to obtain a correct value for the rephasing
invariant $J$ in the quark sector, which is too small in the
conventional 4D USY picture. CP violation in the lepton sector
\cite{Endoh:2002wm} is not measured yet, but it is expected to be
large due to large leptonic mixing.

In addition, the higher dimensional embedding permits large mixing in
the neutrino mass matrix even with hierarchical neutrinos. Note that
in the 4D context, when one imposes USY on the effective light
neutrino mass matrix, neutrinos have to be necessarily
quasi-degenerate in order to achieve large mixing \cite{jugugu}.

\vskip 5pt 

\noindent {\bf{Acknowledgements}:} G.B. acknowledges hospitality at CFTP,
Instituto Superior T\'{e}cnico, Lisbon, where the work was initiated,
and CERN Theory Division (Paid Associates Programme) during the
completion of the work. G.B. also thanks A. Raychaudhuri for comments
on the manuscript. G.C.B. and J.I.S-M. thank CERN Theory Division for
warm hospitality. This work was partially supported by Funda\c c\~ ao
para a Ci\~ encia e a Tecnologia (FCT, Portugal) through projects
PDCT/FP/63914/2005, PDCT/FP/63912/2005, POCTI/FNU/44409/2002 and
CFTP-FCTUNIT 777, which are partially funded through POCTI
(FEDER). The work of G.C.B. was supported by the Alexander von
Humboldt Foundation through a Humboldt Research Award. G.C.B. would
like to thank Andrzej J. Buras for the kind hospitality at TUM.


\begin{thebibliography}{9}
\bibitem{prev-1} Y.~Kawamura, 
%``Triplet-doublet splitting, proton stability and extra dimension,''
Prog.\ Theor.\ Phys.\ \textbf{105} (2001) 999 [arXiv:hep-ph/0012125];
L.~J.~Hall and Y.~Nomura, %``Gauge unification in higher dimensions,'' 
Phys.\ Rev.\ D \textbf{64} (2001) 055003 [arXiv:hep-ph/0103125].

\bibitem{prev-2} L.~Hall, J.~March-Russell, T.~Okui and D.~R.~Smith, 
%``Towards a theory of flavor from orbifold GUTs,'' 
JHEP \textbf{0409} (2004) 026 [arXiv:hep-ph/0108161]; A.~Hebecker and
J.~March-Russell, %``A minimal S(1)/(Z(2) x Z'(2)) orbifold GUT,'' 
Nucl.\ Phys.\ B \textbf{613} (2001) 3 [arXiv:hep-ph/0106166]; Y.~Nomura, 
%``Strongly coupled grand unification in higher dimensions,'' 
Phys.\ Rev.\ D \textbf{65} (2002) 085036 [arXiv:hep-ph/0108170]; A.~Hebecker
and J.~March-Russell, 
%``The flavour hierarchy and see-saw neutrinos from bulk masses in 5d 
%orbifold GUTs,'' 
Phys.\ Lett.\ B \textbf{541} (2002) 338 [arXiv:hep-ph/0205143]; W.~F.~Chang
and J.~N.~Ng, 
%``Neutrino masses in 5D orbifold SU(5) unification models without 
%right-handed singlets,'' 
JHEP \textbf{0310} (2003) 036 [arXiv:hep-ph/0308187]; 
A.~B.~Kobakhidze,
%``Proton stability in TeV-scale GUTs,''
Phys.\ Lett.\  B {\bf 514} (2001) 131 [arXiv:hep-ph/0102323]; 
G.~Altarelli and F.~Feruglio, 
%``SU(5) grand unification in extra dimensions and proton decay,'' 
Phys.\ Lett.\ B \textbf{511} (2001) 257 [arXiv:hep-ph/0102301];
G.~Bhattacharyya and K.~Sridhar, 
%``Testing orbifold models of supersymmetric grand unification,'' 
J.\ Phys.\ G \textbf{29} (2003) 993 [arXiv:hep-ph/0111345].

\bibitem{br} G.~Bhattacharyya and A.~Raychaudhuri, 
%``Remarks on flavour mixings from orbifold compactification,''
J.\ Phys.\ G \textbf{32} (2006) B1 [arXiv:hep-ph/0511276].

\bibitem{usy} G.~C.~Branco, J.~I.~Silva-Marcos and M.~N.~Rebelo, 
%``Universal Strength For Yukawa Couplings,''
Phys.\ Lett.\ B \textbf{237} (1990) 446; 
P.~M.~Fishbane and P.~Kaus,
  %``Pure phase mass matrices,''
Phys.\ Rev.\  D {\bf 49} (1994) 3612; 
G.~C.~Branco and J.~I.~Silva-Marcos, 
%``Predicting V (CKM) with universal strength of Yukawa couplings,''
Phys.\ Lett.\ B \textbf{359} (1995) 166 [arXiv:hep-ph/9507299].

\bibitem{gamma} F.~J.~Botella, G.~C.~Branco, M.~Nebot and M.~N.~Rebelo, 
%``New physics and evidence for a complex CKM,''
Nucl.\ Phys.\ B \textbf{725} (2005) 155 [arXiv:hep-ph/0502133]; The UTfit
collaboration, see the webpage \texttt{http://utfit.roma1.infn.it/}.

%\cite{Endoh:2002wm}
\bibitem{Endoh:2002wm}
  T.~Endoh, S.~Kaneko, S.~K.~Kang, T.~Morozumi and M.~Tanimoto,
  %``CP violation in neutrino oscillation and leptogenesis,''
  Phys.\ Rev.\ Lett.\  {\bf 89} (2002) 231601
  [arXiv:hep-ph/0209020].
  %%CITATION = PRLTA,89,231601;%%


\bibitem{jugugu} G.~C.~Branco, J.~I.~Silva-Marcos and M.~N.~Rebelo, 
%``Universal Strength For Yukawa Couplings and neutrinos, large phases''
Phys.\ Lett.\ B \textbf{428} (1998) 136.
\end{thebibliography}
\end{document}